\documentclass[pra,aps,showpacs]{revtex4}
\usepackage{graphicx}
\usepackage{amsmath}

\begin{document}
\title{Extrinsic electromagnetic chirality in all-photodesigned one-dimensional THz metamaterials}

\author{Carlo Rizza$^{1,2}$, Lorenzo Columbo$^{3,4}$, Massimo Brambilla$^{4,5}$, Franco Prati$^1$, and Alessandro Ciattoni$^{2,}$}
\email{alessandro.ciattoni@spin.cnr.it}
\affiliation{$^1$Dipartimento di Scienza e Alta Tecnologia, Universit\`a degli Studi dell'Insubria, via Valleggio 11, Como, I-22100 Italy}
\affiliation{$^2$Consiglio Nazionale delle Ricerche, CNR-SPIN, via Vetoio 10, L'Aquila, I-67100 Italy}
\affiliation{$^3$Dipartimento di Elettronica e Telecomunicazioni, Politecnico di Torino, Corso Duca degli Abruzzi 24, I-10129 Torino, Italy}
\affiliation{$^4$Consiglio Nazionale delle Ricerche, CNR-IFN, via Amendola 173, I-70126 Bari, Italy}
\affiliation{$^5$Dipartimento Interateneo di Fisica, Universit\`a degli Studi e Politecnico di Bari, via Amendola 173, I-70126 Bari, Italy}

\date{\today}
\begin{abstract}
We suggest that all-photodesigned metamaterials, sub-wavelength custom patterns of photo-excited carriers on a semiconductor, can display an exotic extrinsic electromagnetic chirality in terahertz (THz) frequency range. We consider a photo-induced pattern exhibiting 1D geometrical chirality, i.e. its mirror image can not be superposed onto itself by translations without rotations and, in the long wavelength limit, we evaluate its bianisotropic response. The photo-induced extrinsic chirality turns out to be fully reconfigurable by recasting the optical illumination which supports the photo-excited carriers. The all-photodesigning technique represents a feasible, easy and powerful method for achieving effective matter functionalization and, combined with the chiral asymmetry, it could be the platform for a new generation of reconfigurable devices for THz wave polarization manipulation.
\end{abstract}

\pacs{81.05.Xj, 85.30.-z, 78.20.Ek}
 \maketitle

\section{Introduction}
In the last two decades, metamaterial science has provided novel tools for manipulating electromagnetic radiation \cite{Cai}. In the terahertz (THz) frequency domain, many researchers exploited the potential of metamaterials to fill the so-called THz gap and to conceive efficient optical devices, with particular effort on THz polarization manipulators \cite{Grady,Fan}. Semiconductor based reconfigurable metamaterials show electromagnetic response that can be rapidly modified through photo-doping \cite{Shen,Rizza_3}, application of a bias voltage \cite{Chen_1} and thermal excitation \cite{Gomez_1}. In addition, a spatially modulated optical beam can induce a pattern of photo-carriers which mimics a standard metal-dielectric structure \cite{Okada,Chatzakis,Wang_1,Georgiou}. Considering the homogenized regime (i.e. the light modulation scale generating the photo-carriers pattern is much smaller than the wavelength), we suggested the first example of tunable hyperbolic all-photodesigned THz metamaterial \cite{Rizza_4}. In Ref.\cite{Kam}, using femtosecond laser technology, authors investigated all-photodesigned transient metamaterials allowing the manipulation of THz waveforms polarization with subcycle switch-on times. Recently, Mezzapesa et al. \cite{Mezzapesa} devised an all-photodesigned metamaterial reflector and they exploited it to control the emission properties of a THz QCL. All-photodesigning technique provides ultra-fast reconfigurability and it is a type of grey-scale lithography \cite{Kam}. Furthermore, in the quasi-homogenized regime, by tailoring the geometry of the photoinduced carrier profile, one can achieve a strong electromagnetic chiral response. In Ref.\cite{Kanda}, the authors experimentally realized all-photodesigned gammadion-type metamaterials showing a highly tunable optical activity. Artificial electromagnetic chirality and bi-anisotropy (due to the magneto-electric coupling) have attracted much attention in metamaterial research and they support several relevant effects such as, for example, negative refraction \cite{NEG},  giant optical activity, asymmetric transmission \cite{asim}. It is worth noting that the reciprocal bianisotropy response can also be achieved in a metamaterial whose basic meta-atoms do not show the standard 2D-3D geometrical chiral asymmetry. This peculiar electromagnetic response is termed as \textit{extrinsic} electromagnetic chirality \cite{Plum} and it can yield giant linear  \cite{Yoko,Leon} and nonlinear \cite{Belardini} circular dichroism. Moreover, in Ref.\cite{Rizza_2}, the authors introduced the concept of 1D geometrical chirality (namely the mirror image of the considered pattern can not be superposed onto itself by translations without rotations) and they proved that a 1D multi-layered structure, without intrinsic chiral inclusions, can support strong extrinsic electromagnetic chirality.

In this paper, we introduce all-photodesigned THz metamaterials exhibiting 1D chirality. From full-wave simulations, we numerically retrieve the dielectric and chiral tensor of the considered metamaterial. By fully exploiting the all-photodesign technique, we show that the extrinsic chirality is highly reconfigurable by varying the intensity and the chiral degree of the light pattern generating photo-excited carriers. We support the numerical evidences through a simplified approach providing an analytical description of the non-local effective medium.

\section{THz metamaterial photodesigning}

In Fig.1, we report the sketch of an all-photodesigned 1D chiral THz metamaterial and its corresponding waves scattering geometry. An infrared (IR) plane wave, modulated along the $x$-axis by a spatial light modulator (SLM), normally impinges onto the vacuum-semiconductor interface. The light beam produces a photo-generated charge carrier pattern which spatially modulates the local plasma frequency thus yielding a spatially modulated dielectric permittivity profile at THz frequencies \cite{Rizza_4}.
\begin{figure}
\centering
\includegraphics[width=0.5\textwidth]{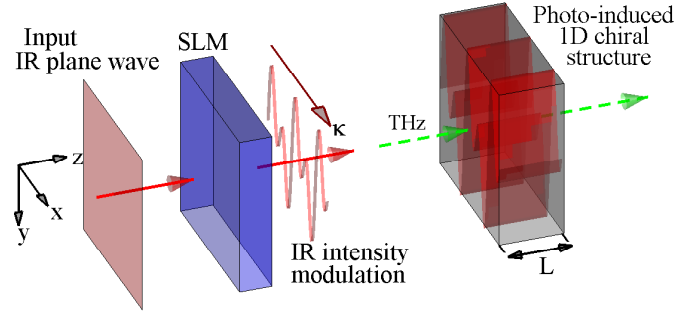}
\caption{(Color online) Sketch of the all-photodesigned 1D metamaterial and waves scattering. An infrared (IR) plane wave is modulated by a spatial light modulator (SLM) and it generates a photo-induced 1D chiral structure (whose strength and orientation is identified by 1D chiral vector $\boldsymbol{\kappa}$) within the bulk of a semiconductor slab of thickness $\mathrm{L}$. THz waves (dashed arrows) normally impinge onto the slab.}
\end{figure}
Here, we consider the situation where the modulated beam intensity and the corresponding "microscopic" dielectric permittivity $\varepsilon^{(THz)}$ are 1D periodic functions showing 1D chiral asymmetry \cite{Rizza_2} and whose period $\Lambda$ is much smaller than the THz wavelength $\lambda^{(THz)}$. In the long wavelength limit ($\eta= \Lambda / \lambda^{(THz)} \ll 1$), the effective constitutive relations are
\begin{eqnarray}
\label{cost-rel}
{ D}_x &=& \varepsilon_0 \left [ \varepsilon_{\parallel} { E}_x + \frac{\kappa}{k_0^{(THz)}} \left( \partial_y { E}_y +\partial_z { E}_z \right)  \right],  \nonumber \\
 { D}_y &=& \varepsilon_0 \left( \varepsilon_{\perp} { E}_y - \frac{\kappa}{k_0^{(THz)}} \partial_y { E}_x \right),    \nonumber \\
 { D}_z &=& \varepsilon_0\left( \varepsilon_{\perp} {\ E}_z - \frac{\kappa}{k_0^{(THz)}}  \partial_z { E}_x \right),
\end{eqnarray}
where $k_0^{(THz)}=2\pi/\lambda^{(THz)} = \omega^{(THz)}/c$ and $\varepsilon_{\parallel}$,$\varepsilon_{\perp}$ and $\kappa$ are the three parameters describing the effective electromagnetic response. The parameter $\kappa$ measures the effects of the first order spatial dispersion and its amplitude is proportional to $\eta$ \cite{Rizza_2}. As it is well-known, first order spatial dispersion is physically equivalent to reciprocal bianisotropy \cite{Serdy} and such equivalence is expressed by the Serdyukov-Fedorov transformation \cite{Serdy}, i.e. ${\bf { D}}'={\bf { D}} - \nabla \times {\bf Q}$, ${\bf { H}}'={\bf { H}} + i \omega^{(THz)} {\bf Q}$ with ${\bf Q}=-\varepsilon_0 \kappa/ k_0^{(THz)} \left( { E}_z {\hat { \bf e}_y}-{ E}_y {\hat { \bf e}_z} \right)$. Here, we use such transformation to investigate the chiral properties of the photoinduced metamaterial and we get
\begin{equation}
\label{const-eq}
{\bf  D}' = \varepsilon_0 \varepsilon^{(eff)} {\bf E} - \frac{i}{c} \boldsymbol{\kappa} \times {\bf H}', \quad
{\bf  B} =  -\frac{i}{c} \boldsymbol{\kappa} \times {\bf  E} + \mu_0 {\bf  H}',
\end{equation}
where $\varepsilon^{(eff)}= \textrm{diag}(\varepsilon_{\parallel},\varepsilon_{\perp}',\varepsilon_{\perp}')$ ($\varepsilon_{\perp} ' = \varepsilon_{\perp} + \kappa^2$). In Eqs.(\ref{const-eq}), we have introduced the 1D chiral vector $\boldsymbol{\kappa}= \kappa \hat{\bf e}_x$ which does not vanish only if the photo-induced pattern has 1D chiral asymmetry along the $x$-axis \cite{Ciattoni_1}.

In order to investigate the chiral electromagnetic response of 1D photoinduced metamaterials, we have performed 2D full-wave simulations  \cite{comsol} where Maxwell's equations for both the $1$+$1$D monochromatic IR beam and the THz field are coupled with the rate equation for the carriers dynamics. We have focused on a GaAs slab and, by using the microscopic approach reported in Ref.\cite{Tissoni}, we have evaluated the density-dependent optical semiconductor dielectric constant $\varepsilon^{(IR)}(N)$ (see Supplementary Material). In the steady state, the $2$D spatial dynamics of the carrier density $N(x,z)$ is described by the rate equation
\begin{equation}
\label{rate-eq}
D (\partial_x^2+\partial_z^2) N - \frac{N}{\tau(N)} + \frac{\varepsilon_0 }{2 \hbar } Im \left[ \varepsilon^{(IR)}(N) \right] |{\bf E}^{(IR)}|^2=0,
\end{equation}
where ${\bf E}^{(IR)}(x,z)$ is the electric field at the considered optical frequency inside the semiconductor slab, $\tau(N)$ is the density dependent electron-hole recombination time and $D$ is the ambipolar diffusion coefficient ($\hbar$ is the reduced Planck's constant) \cite{Garmire}. Here, the density dependent electron-hole recombination time is given by $1/\tau=A N + B N^2 + C N^3$, where $A$ and $B$ are the non-radiative and radiative recombination rates, respectively, while the term proportional to $N^3$ describes Auger recombinations. The THz dielectric response is described by the Drude model
\begin{eqnarray}
\label{perm_THz}
\varepsilon^{(THz)} &= &\varepsilon_b + g N,
\end{eqnarray}
where $g = - e^2/ \left(\varepsilon_0 m^* \right) \left[\omega^{(THz)} \left( \omega^{(THz)}+i \gamma_D \right) \right]^{-1}$, $\varepsilon_b$ is the THz background dielectric constant in the absence of the optical beam, $-e$ is the electron charge unit, $m^*$ is the electron effective mass and $\gamma_D$ is the free electron relaxation rate.

\section{Photoinduced extrinsic chirality}

As a theoretical benchmark to discuss the properties of photo-induce chiral phenomena, we focus on the case where the incident IR illumination is given by
\begin{equation} \label{c_s}
I_{in}(x)= I_0 \left\{ 1+ \frac{1}{2} \left[ \cos{\left(\frac{2 \pi}{\Lambda} x \right)} +\chi \sin\left(\frac{4 \pi}{\Lambda} x\right) \right]   \right\},
\end{equation}
where $I_0$ is the average intensity and $\chi$ is a parameter controlling the illumination chiral symmetry and, consequently, the photo-induced grating one (the allowed values of $\chi$ are those for which $I_{in}$ is positive). It is worth noting that, for $\chi \neq 0$, the considered intensity profile $I_{in}(x)$ exhibits geometrical 1D chirality, namely its mirror image ($I_{in}(-x)$) can not be superposed onto itself by 1D translations ($I_{in}(x+x_0) \neq I_{in}(-x)$ for all $x_0$); whereas, for $\chi=0$, the intensity profile is an even function and it is achiral.

In the considered numerical examples, we have set $\lambda^{(IR)}=870$ nm for the IR wavelength, $\Lambda=10$ $\mu$m for the grating period, $L=2.5$ $\mu$m for the semiconductor thickness and we have used typical values for GaAs paramaters: $A=10^8$ s$^{-1}$ \cite{Tissoni}, $B=7.2 \cdot 10^{-16}$ m$^3$/s, $C=10^{-42}$ m$^6$/s \cite{Varr}, $D=A L_D^2$ where the diffusion length is $L_D=2$ $\mu$m, $m^*=0.067 \quad m_0$ ($m_0$ is the electron mass) and $\gamma_D=3.09$ THz \cite{Black}. In addition, we have modeled the background dielectric constant $\varepsilon_b$ (appearing in Eq.(\ref{perm_THz})) with the expression $\varepsilon_b=\varepsilon_1 - \left[\varepsilon_1 (\omega_L^2-\omega_T^2)\right]/ \left[\omega^{(THz)}  \left(\omega^{(THz)} + i \gamma \right) -\omega_T^2 \right]$, where $\varepsilon_1=11$, $\omega_L=55.06$ THz, $\omega_T=50.65$ THz and $\gamma=0.45$ THz \cite{Palik}. The IR beam is a transverse electric (TE) wave, i.e. ${\bf E}_{in}^{(IR)}= E_{in}^{(IR)}(x,z) \exp {(-i\omega^{(IR)} t)} \hat{\bf e}_y$ (being $\omega^{(IR)}$ the angular IR frequency) whose incident intensity is $I_{in}(x)$ of Eq.(\ref{c_s}).

The incident THz fields are monochromatic plane waves normally impinging onto the GaAs slab (linearly polarized in the $x$-$y$ plane). The semiconductor slab has been placed between two vacuum layers. Along the $x$-axis, we have considered periodic conditions for IR, THz and carrier density; whereas, along the $z$-axis, we have set scattering boundary and matched boundary conditions for IR and THz waves, respectively. We have required the charge current to vanish (i.e. $\partial_z N=0$) at the facets of the semiconductor slab.
\begin{figure}
\centering
\includegraphics[width=0.5\textwidth]{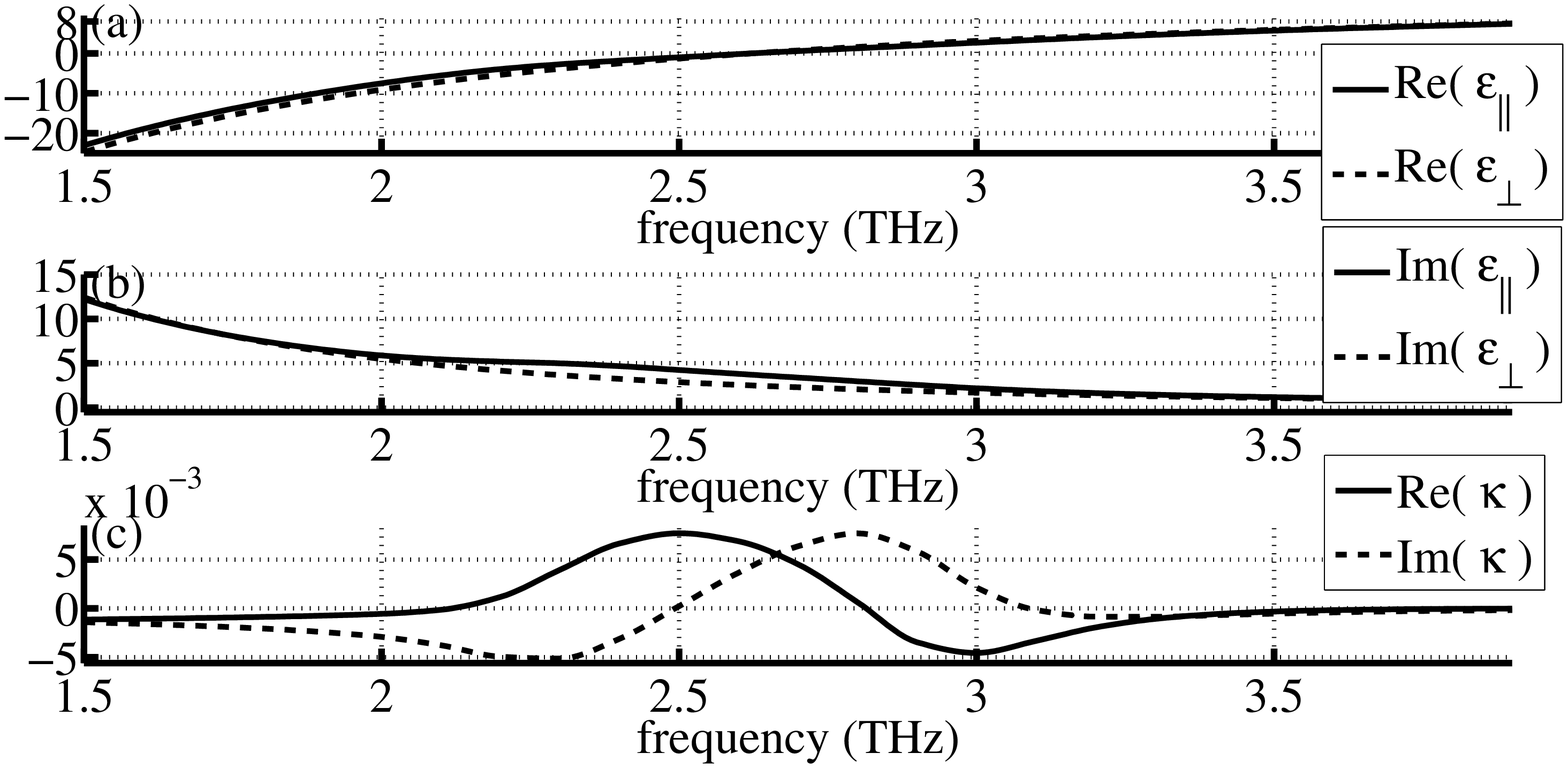}
\caption{THz frequency dependence of the real (a) and the imaginary (b) parts of the dielectric permittivities ($ \varepsilon_{\parallel} $, $ \varepsilon_{\perp}$). (c) THz frequency dependence of the real and imaginary parts of the chiral electromagnetic parameter $\kappa$. Here $I_0=1.85$ kW/cm$^2$ and $\chi=1$.}
\end{figure}
We have focused on THz frequencies $\nu^{(THz)}$ in the range $1.5 - 4$ THz and accordingly, since the period to wavelength ratio $\eta$ is smaller than $0.13$, we have assumed the THz electromagnetic response of the photo-induced grating to be quasi-homogenized and to be given by the effective medium description of Eqs.(\ref{cost-rel}). Strictly, Eqs.(\ref{cost-rel}) are valid only for a one-dimensional configuration where the microscopic permittivity only depends on $x$. In the considered situations, the microscopic dielectric profile retains a dependence from $z$ which is inherited from the spatial pattern of the IR illumination within the slab. However, the chosen diffusion length $L_D$ (characterizing the spatial spreading of the photo-carrier distribution) is much larger than the IR wavelength so that the profile of the photo-induced carrier density $N$ is largely smoothed by carrier diffusion to the point that $N$, and hence $\varepsilon^{(THz)}$ of Eq.(\ref{perm_THz}), can be regarded as uniform along the $z$-axis within the slab (see Supplementary Material) \cite{Rizza_1}.

In order to numerically retrieve the effective parameters appearing in Eqs.(\ref{cost-rel}) from the full-wave simulations we have averaged the fields along the $x$-axis and subsequently integrated such averages along the slab thickness. Therefore, after integrating both sides of Eqs.(\ref{cost-rel}) along the $z$-axis inside the slab, we have retrieved $\varepsilon_{\parallel}$, $\varepsilon_{\perp}$ and $\kappa$ by solving the three algebraic equations thus obtained.

The general features of the retrieved THz effective parameters are reported in Fig.2 where we plot their characteristic spectral profiles for an IR illumination with $I_0=1.85$ kW/cm$^2$ and $\chi=1$. In Fig.2(a) and 2(b) we plot the real and imaginary parts of the effective dielectric permittivities $\varepsilon_{\parallel}$, $\varepsilon_{\perp}$; whereas, in panel (c), we report the real and the imaginary part of the chiral electromagnetic parameter $\kappa$. The homogenized response is slightly birefringent ($ \varepsilon_{\parallel}  \simeq  \varepsilon_{\perp} $) and it shows extrinsic electromagnetic chirality ($ \kappa \neq 0$). In the considered situation, we have chosen the IR intensity profile in such a way that the effective permittivity has a zero crossing-point within the considered frequency range (i.e. $Re(\varepsilon_{\perp})=0$ for $\nu^{(THz)}_0=2.7$ THz). Note that the profile of $\kappa$ shows a distinct and broad resonance-like shape in a spectral region surrounding the crossing point $\nu^{(THz)}_0$ which is a consequence of the competition between semiconductor dispersion (Drude model of Eq.(\ref{perm_THz})) and averaging features yielding electromagnetic chirality, as will become clearer in the following. Bearing in mind that standard SRR-based THz metamaterials  are characterized by a narrow-band frequency response, the considered broadband magneto-electric response can be a relevant feature for different applications \cite{broad}.
\begin{figure}
\centering
\includegraphics[width=0.5\textwidth]{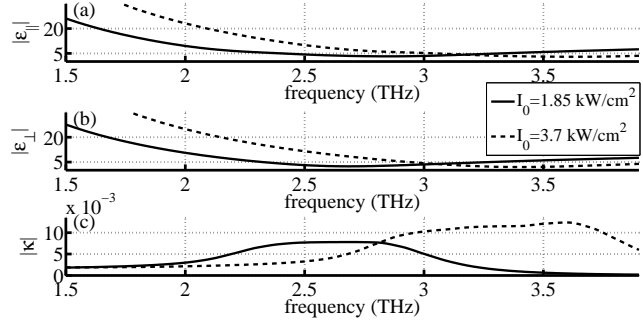}
\caption{$|\varepsilon_{\parallel}|$ (a), $|\varepsilon_{\perp}|$ (b) and $|\kappa|$ (c) as functions of the THz frequency for $I_0=1.85$ kW/cm$^2$ (solid lines) and $I_0=3.7$ kW/cm$^2$ (dashed lines). Here $\chi = 1$.}
\end{figure}
Compared with active composite metamaterials (see e.g. \cite{Kan}) where the geometry design is structural and thus rigid, all-photodesigned metamaterials offer improved flexibility. The high tunability of all-photodesigned THz metamaterials is due to two significant reasons: (i) the incident IR profile can be easily and rapidly modified at will (the dielectric permittivity is written, erased and rewritten even on ultrafast time scales \cite{Kam}), (ii) the possibility to achieve a continuous spatial variation of the microscopic permittivity (grey-lithography) greatly increases the possibilities for the effective electromagnetic behavior. In our examples, the freedom to independently set the IR intensity $I_0$ and the chirality degree $\chi$ of the considered intensity profile allow to investigate the tunability features of all-photodesigned THz metamaterials.

In Fig.3, we plot the absolute value of $\varepsilon_{\parallel}$, $\varepsilon_{\perp}$ and $\kappa$ (panels (a), (b) and (c), respectively) for two different optical intensities, namely $I_0=1.85$ kW/cm$^2$, $I_0=3.7$ kW/cm$^2$ and $\chi=1$. For both intensities, the permittivity absolute value has a minimum at the zero-crossing point. Accordingly, the profile of $|\kappa|$ shows a broad hump, due to the broad resonance of $Re(\kappa)$ and $Im(\kappa)$, which for both intensities is located around the corresponding permittivity crossing-point. Note that, remarkably, the higher IR intensity provides the stronger electromagnetic chirality. In Fig.4, we plot the absolute value of $\varepsilon_{\parallel}$, $\varepsilon_{\perp}$ and $\kappa$ (panels (a), (b) and (c), respectively) for three different chirality degrees, namely $\chi =0, 0.5, 1$ for the IR intensity $I_0=1.85$ kW/cm$^2$. While $|\varepsilon_{\parallel}|$ and $|\varepsilon_{\perp}|$ are almost independent of $\chi$, $|\kappa|$ is effectively proportional to $\chi$ and it vanishes for $\chi=0$, as it must be, since in this case the IR intensity profile is achiral. In panel (d) of Fig.4, for the same values of $\chi$ and $I_0$ as above, we plot $|\langle E_z \rangle|$, where $\langle f(z) \rangle =\frac{1}{L} \int_0^L dz f(z)$ labels the spatial average over the slab thickness and $E_z$ is the $z$-component of the THz electric field. Note that the fact that $|\langle E_z \rangle|$ is not vanishing is a key signature of the photoinduced chiral response. In fact the launched THz plane-wave normally impinging at the entrance slab facet has vanishing longitudinal component so that the matching conditions across such facet requires the vanishing of $D_z$ within the slab and the third of Eqs.(\ref{cost-rel}) correspondingly implies that $E_z \neq 0$ if $\kappa \neq 0$.

\begin{figure}
\centering
\includegraphics[width=0.5\textwidth]{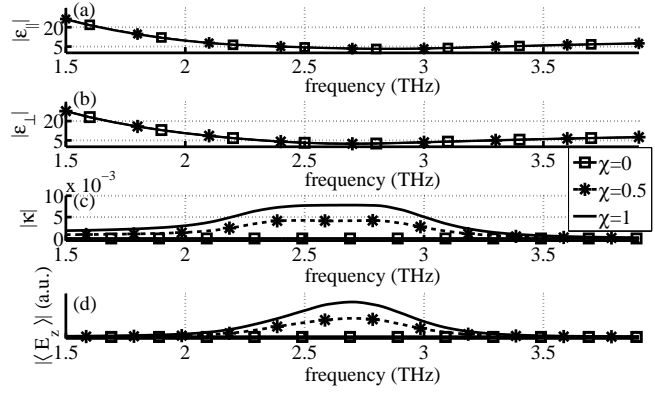}
\caption{THz frequency dependence of $|\varepsilon_{\parallel}|$ (a), $|\varepsilon_{\perp}|$ (b), $|\kappa|$ (c) and $|\langle E_z \rangle |$ (d), for $\chi=0$ (squares and solid lines), $\chi=0.5$ (stars and dashed lines), $\chi=1$ (solid line) for $I_0=1.85$ kW/cm$^2$.}
\end{figure}

\section{Discussion}

In order to physically grasp the basic phenomenological features of the all-photoinduced metamaterials we exploit the approach of Ref.\cite{Rizza_L} which, for a metamaterial with low dielectric contrast, provides an analytical description of the effective medium bianisotropy. Such approach is suitable to describe the examples considered in this paper since the two permittivities $\varepsilon_{\parallel}$ and $\varepsilon_{\perp}$ can be slightly different (see Fig.2) only if the absolute value of the photoinduced dielectric modulation $|\Delta \varepsilon^{(THz)}|  =| \varepsilon^{(THz)} - \bar \varepsilon|$ is much smaller than $|\bar \varepsilon|$ ($\bar \varepsilon$ is the spatial average of the THz permittivity). Since we are interested in understanding the main features of the obtained numerical results, to simplify the treatment we do not consider IR propagation within the thin slab ($|E^{(IR)}|^2=2 I_{in}/(\varepsilon_0 c)$) and we neglect both nonlinear contributions and the $z$-dependence of $N$ in Eq.(\ref{rate-eq}), thus obtaining for the photoinduced carried density
\begin{equation} \label{NN}
N = \beta I_0 \left\{ {1 + d_1 \cos \left( {\frac{{2\pi }}{\Lambda }x} \right) + \chi d_2 \sin \left( {\frac{{4\pi }}{\Lambda }x} \right)} \right\},
\end{equation}
where $\beta  = \frac{{{\mathop{\rm Im}\nolimits} \left( {\varepsilon^{(IR)}(0) } \right)}}{{A \hbar c}}$, $d_1  = \frac{1}{2}\left[ {1 + \left( {2\pi \frac{{L_D }}{\Lambda }} \right)^2 } \right]^{ - 1}$, $d_2  = \frac{1}{2}\left[ {1 + \left( {4\pi \frac{{L_D }}{\Lambda }} \right)^2 } \right]^{ - 1}$ and $Im(\varepsilon^{(IR)}(0))=0.14$. After substituting Eq.(\ref{NN}) into Eq.(\ref{perm_THz}) and following the approach of Ref.\cite{Rizza_L} (see Supplementary Material), we obtain
\begin{eqnarray} \label{approx}
\varepsilon _\parallel   &=&
 \bar \varepsilon - \frac{1}{4} \left[ d_1^2 + \left( d_2 \chi  \right)^2  \right]  \frac{\left(g \beta  I_0 \right)^2 }{\bar \varepsilon },
\quad \quad \varepsilon _ \perp   = \bar \varepsilon, \nonumber \\
\kappa  &=& \eta \chi  \frac{3  d_1^2 d_2}{8} \frac{\left({ {g\beta  I_0 } }\right)^3}{{\bar \varepsilon}^2},
\end{eqnarray}
where $\bar \varepsilon  = \varepsilon _b  + g \beta I_0$. From Eqs.(\ref{approx}) it is evident that the birefringence $|\varepsilon _\parallel - \varepsilon_{\perp}|$ and the chiral parameter $\kappa$ are proportional to $(\bar \varepsilon)^{-1}$ and to $(\bar \varepsilon)^{-2}$, respectively. Accordingly, in the spectral region around the permittivity crossing point, this proves both the slight discrepancies between $\varepsilon_{\parallel}$ and $\varepsilon_{\perp}$ in panel (a) and (b) of Fig.2 and the resonant-like shape of $\kappa$ in panel (c) of Fig.2. Therefore in the low-contrast regime, the smaller values of the average permittivity entail a larger electromagnetic chirality. Physically this is due to the fact that, when the dielectric contrast is very low, the difference between the dielectric profile and its mirror image is magnified for lower values of the average $\bar \varepsilon$. Note that the dependence of $\bar \varepsilon$ on $I_0$ agrees with the metamaterial tunability features reported in Fig.3. Finally, the expression of $\kappa$ in the third of Eqs.(\ref{approx}) suggests different ways for increasing the value of such chiral parameter. This can be done, for example, by shaping the IR illumination pattern with higher chiral degree $\chi$ (as shown in panel (c) of Fig.4) and with higher average intensity $I_0$ or by designing the semiconductor bulk with smaller diffusion length $L_d$ and higher IR absorption coefficient $\beta$.

\section{Conclusions}

In conclusion, we have introduced all-photodesigned THz metamaterials displaying 1D chirality. We have investigated the tunability of the extrinsic electromagnetic chiral response of a 1D photo-induced structure, where the magneto-electric coupling is due the 1D chiral asymmetry of the photo-carrier density profile. Such 1D photo-induced structures show the same kind of electromagnetic chiral response exhibited by chiral layered media which support optical activity (see Ref.\cite{Rizza_2}) and therefore they can be exploited to manipulate THz radiation. The strength of electromagnetic chirality can in principle be increased by choosing highly asymmetric infrared illumination profiles, by extending our approach to photo-induced 2D structures or by resorting to femto-second laser technology to enhance the photo-induced dielectric modulation depth \cite{Kam}. In view of the high spatial and temporal reconfigurability shown by this kind of THz metamaterials, all-photodesigned chirality holds the potential to conceive a novel class of active optical devices such as, for examples, circular polarization beam splitters and polarization spectral filters.

A. C. and C. R. thank the U.S. Army International Technology Center Atlantic for financial support (Grant No. W911NF-14-1-0315).

\section*{Supplementary Material}

\subsection{Light-matter coupling in a bulk semiconductor}

Following the approach in \cite{Tissoni}, we report here the First-Principles calculation of the semiconductor optical susceptibility $\chi(N)$ used to evaluate at the frequency $\omega^{(IR)}=2168$ THz ($\lambda^{(IR)}=870$ nm) the density-dependent dielectric constant $\varepsilon^{(IR)}(N)$ in Eq.(3) of the main manuscript. We will focus on a GaAs bulk semiconductor medium. \\

In the free carriers, quasi-equilibrium and the two-bands approximations the expression of the complex susceptibility which describes the radiation-matter interaction reads \cite{haug,koch}
\begin{equation}
\chi (N)=\frac{-i}{\hbar \varepsilon _{0}V}\sum_{\mathbf{k}}\frac{%
\mu _{\mathbf{k}}^{2}(f_{e\mathbf{k}}(N)+f_{h\mathbf{k}}(N)-1)}{\gamma_{p}+i(\omega_{\mathbf{k}}-\omega)},  \label{chical}
\end{equation}
where $\mathbf{k}$ is the carrier momentum, $\mu_{\mathbf{k}}$ is the dipole matrix element associated with an optical transition,
$V$ is the active volume, $\gamma_{p}$ is the polarization decay rate ($\sim 3.4 \times10^{13} \, {\rm s}^{-1}$ for GaAs), $N$ is the total carrier density, $\omega$ is the injected field frequency and
\begin{equation}
\hbar \omega_{\mathbf{k}}=E_g+\epsilon_{e}+\epsilon_{h}=E_g+ \frac{\hbar^{2}k^{2}}{2m_{e}}+ \frac{\hbar^{2}k^{2}}{2m_{h}}
\end{equation}
is the transition energy at the carrier momentum $\mathbf{k}$, being $E_g$ the energy gap ($E_g=1.424$ eV for GaAs) and $m_{e,h}$ the effective mass of the electrons in the conduction band and of the holes in the valence band ($m_{e}=0.067m_{0}$, $m_{h}=0.57m_{0}$ for GaAs \cite{Bouarissa} where $m_{0}$ is the free electron mass). We set the zeros of $\epsilon _{e}$ and $\epsilon _{h}$ equal to the bottom of the conduction band and the top of the valence band respectively. The Fermi-Dirac distributions in Eq.(\ref{chical}) are given by
\begin{equation}
f_{e\mathbf{k}} (N)=\frac{1}{\exp (\frac{\epsilon _{e}-\mu _{e}}{k_{b}T})+1}, \quad
f_{h\mathbf{k}} (N)=\frac{1}{\exp (\frac{\epsilon _{h}-\mu _{h}}{k_{b}T})+1},
\end{equation}
where $\mu _{e}(N)$ and $\mu _{h}(N)$ are the electron and holes quasi-chemical potential, $k_{b}$ is the Boltzmann constant and $T$ is the device temperature.

Using a series representation for the Fermi-Dirac functions and the Pad\`{e} approximation, it is possible to derive the following analytical approximations for the chemical potential valid in the bulk case \cite{haug, koch}
\begin{eqnarray}
\mu _{e} &=&(\ln N_{e}+k_{1}\ln (k_{2}N_{e}+1)+k_{3}N_{e})k_{b}T, \nonumber \\
\mu _{h} &=&(\ln N_{h}+k_{1}\ln (k_{2}N_{h}+1)+k_{3}N_{h})k_{b}T,
\end{eqnarray}
where  $k_{1}=4.8966851$, $k_{2}=0.04496457$, $k_{3}=0.1333760$ and $N_{e}$ and $N_{h}$ are the normalized total carrier densities
given by
\begin{equation}
N_{e} =4N\left( \frac{\pi \hbar ^{2}}{2k_{b}Tm_{e}}\right) ^{3/2}, \quad
N_{h} =4N\left( \frac{\pi \hbar ^{2}}{2k_{b}Tm_{h}}\right) ^{3/2}.
\end{equation}
Moreover, for the square modulus of the electric dipole element $\mu _{\mathbf{k}}$ we assume the expression \cite{Boer, Corzine, Bava}:
\begin{equation}
\mu _{\mathbf{k}}^{2}=\frac{E_g(E_g+\Delta_{s})}{4\left(E_g+\frac{2}{3}\Delta_{s}\right)}\left( \frac{1}{m_{e}}-\frac{1}{m_{0}}\right)\left(  \frac{e\hbar}{\hbar \omega_{\mathbf{k}}}\right)^{2}
\end{equation}
where $\Delta_{s}$ is the spin-orbit energy splitting ($\Delta_{s}=0.34$ eV for GaAs).

\noindent In order to phenomenologically correct the overestimation of the homogeneous broadening effects of the Lorentzian slowly decaying tails (leading to an excessive absorption at photon frequencies below the band gap) we have considered an exponential dependence on $k$ in the polarization decay rate $\gamma_{k}$ (Urbach tail correction) \cite{haug,Boer}
\begin{equation}
\gamma_{p} \longrightarrow \gamma_{k}=\frac{2 \gamma_{p}}{\exp(\frac{\hbar \omega_{\mathbf{k}}-\hbar \omega}{E_{0}})+1},
\end{equation}
where $E_{0}$ is an empirical parameter linked to the LO phonon energy ( $E_{0} \simeq 0.032$ eV for GaAs). The value of the energy gap $E_g$ is corrected to account for many-body effects by extrapolating the band gap reduction $\Delta_{g}$ as a function of the carrier density from a semianalytical curve well known in literature \cite{koch}
\begin{equation}
\Delta_{g}=E_{R}\left( -4.29115 e^{-(r_s+0.09274)/1.3888}-27.14073e^{-(r_s+0.09274)/0.27683} \right)
\end{equation}
where $r_s=\left(\frac{3}{4\pi N a_{0}^{3}}\right)^{1/3}$ is the scaled interparticle distance, being $a_{0}=\frac{4\pi\epsilon_{0}n_{0}^{2}\hbar^{2}}{e^{2}m_{r}}$ the exciton Bohr radius and $E_{R}=\frac{e^{2}}{8\pi\epsilon_{0}n_{0}^{2}a_{0}}$ the $3D$ exciton Rydberg energy. In the expression of $a_{0}$ the constant $n_{0}$ represents the background refractive index ($n_{0}=3.6$ for GaAs) and $m_{r}$ is the reduced electron-hole mass defined as
$\frac{1}{m_{r}}=\frac{1}{m_{e}}+\frac{1}{m_{h}}$. We further observe that the semi-analytical curve giving $\Delta_{g}/E_{R}$ as a function of the interparticle distance $r_s$ is almost the same for several semiconductor materials \cite{Klingshirn}. Finally for a bulk medium, assuming an essentially continuous distribution of the states, we can replace the summation $\frac{1}{V}\sum\limits_{\mathbf{k}}$ with the integral $\frac{2}{\left( 2\pi \right) ^{3}} \int\limits_{0}^{\infty }4\pi k^{2}dk$. \\

\begin{figure}[htb]
\centering
\includegraphics[width=0.4\textwidth]{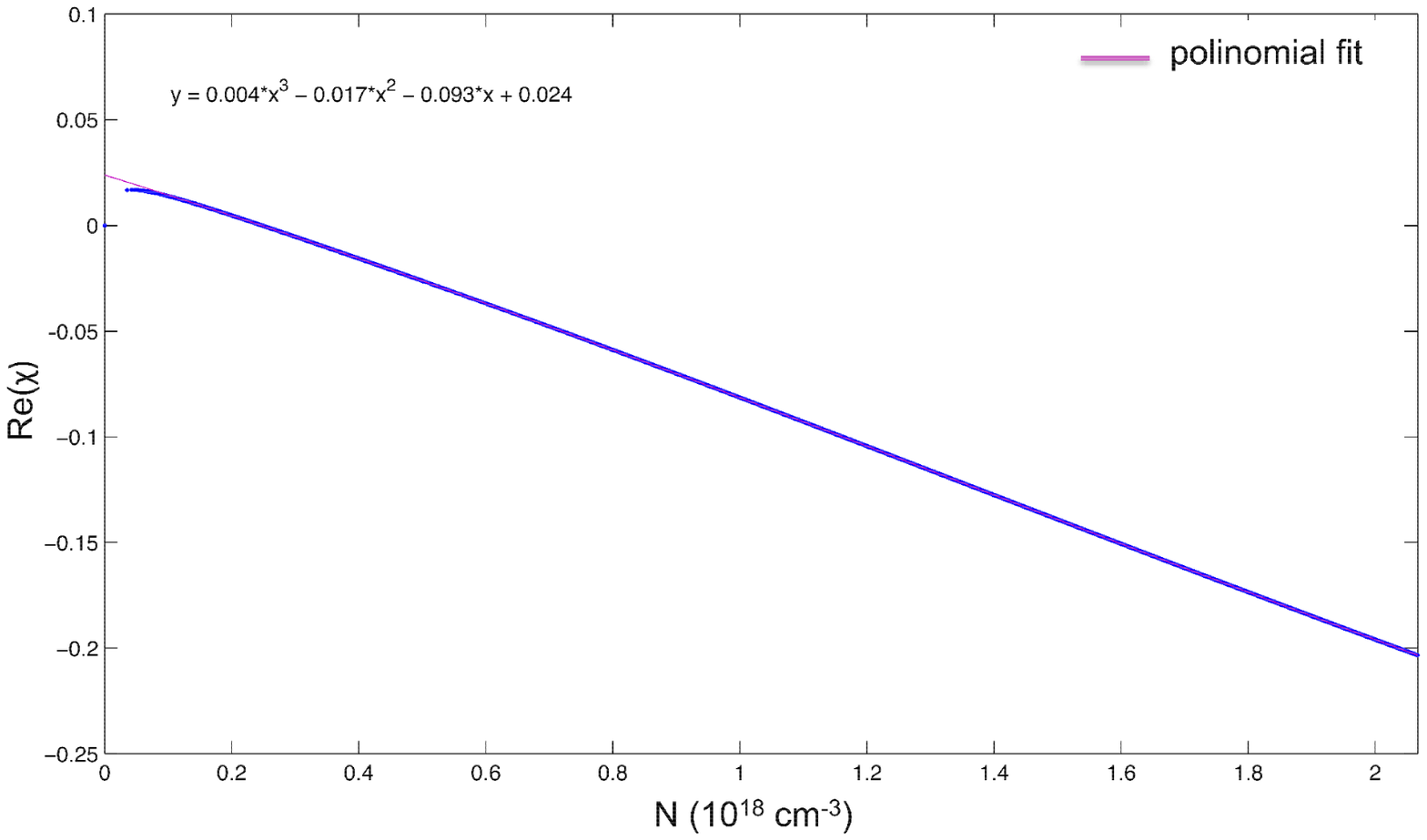}
\includegraphics[width=0.4\textwidth]{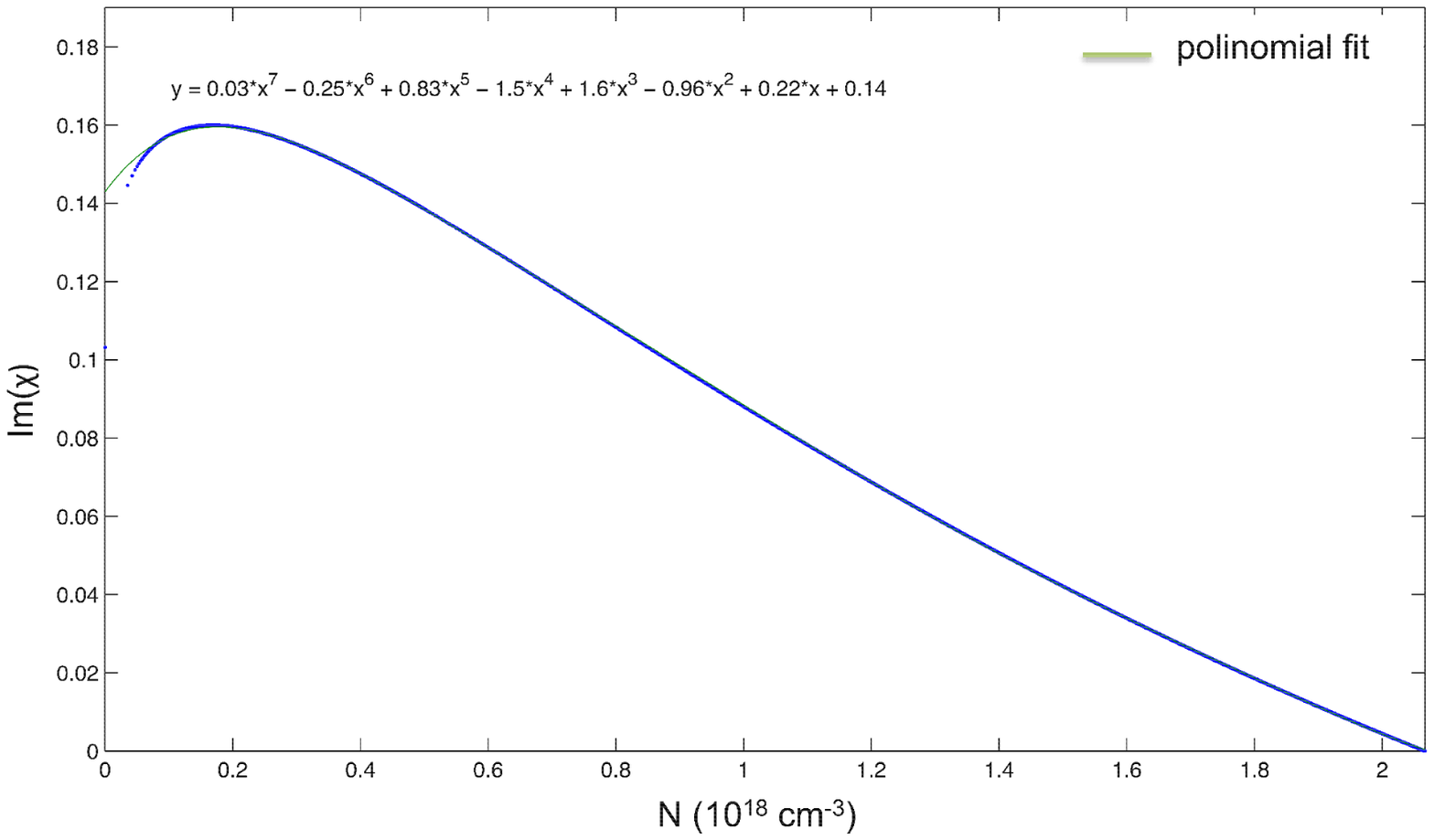}
\caption{Real and imaginary parts of the susceptibility $\chi$ for bulk GaAs versus the carrier density $N$ for $\omega=\omega^{(IR)}=2168$ THz.} \label{fig_chi}
\end{figure}

In Fig.\ref{fig_chi} we plot the real and imaginary part of $\chi$ versus $N$ for $\omega=\omega^{(IR)}=2168$ THz. A polynomial fitting gives for the carriers density-dependence of the real and imaginary part of $\chi$ respectively
\begin{eqnarray}
Re \left[\chi(N) \right] &=& 0.004(N/N_{0})^{3}-0.017(N/N_{0})^{2}-0.093(N/N_{0})+0.024, \nonumber \\
Im\left[\chi(N)\right]&=&0.03(N/N_{0})^{7}-0.25(N/N_{0})^{6}+0.83(N/N_{0})^{5}-1.5(N/N_{0})^{4}  \nonumber \\
&+&1.6(N/N_{0})^{3}-0.96(N/N_{0})^{2}+0.22(N/N_{0})+0.14
\end{eqnarray}
where $N_{0}=10^{18} cm^{-3}$, which are used in turn to evaluate the dielectric constant $\varepsilon^{(IR)}$ according to the formula
\begin{equation}
\varepsilon^{(IR)}(N)=n_{0}^{2}+{\chi}(N).
\end{equation}

\subsection{Effect of carrier diffusion on the photoinduced THz dielectric profile}

\begin{figure}[h]
\centering
\includegraphics[width=0.5\textwidth]{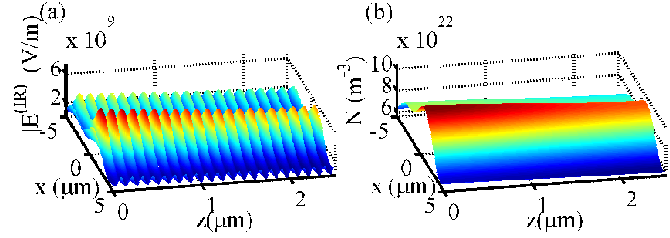}
\caption{(Color on-line) (a) Infrared electric field absolute value $|E^{(IR)}|$ and (b) photo-carrier density $N$ for $\chi=1$ and $I_0=1.85$ kW/cm$^2$.} \label{N_E}
\end{figure}

The basic assumptions on the photoinduced THz dielectric profile $\varepsilon^{(THz)}$ in the manuscript are that 1) it does not depend on $z$ and 2) it reproduces the modulation of the infrared illumination along the $x$-axis. Such assumptions can be made consistent by choosing the diffusion length $L_D$ satisfying the constraint $\lambda^{(IR)} \ll L_D \ll \Lambda$ where $\lambda^{(IR)}$ and $\Lambda$ are the radiation wavelength and the infrared pattern period. For the numerical parameters used in the main manuscript, and for the IR illumination of Eq.(5) of the main manuscript with $\chi=1$ and $I_0=1.85$ kW/cm$^2$ we have evaluated the photoinduced carrier distribution and in Fig.\ref{N_E}, we report both $|E^{(IR)}(x,z)|$ (a) and $N(x,z)$ (b). Note that longitudinal dynamics of the IR field is dominated by the interference between the forward and backward components, whereas the photo-carrier density $N$ is almost independent on $z$. On the other hand the transverse modulation of the IR field is encoded in the transverse profile of $N$.

\subsection{A First-Principles Homogenization approach for one-dimensional photo-induced THz metamaterials}

In this section we report the derivation and the main results of a homogenization approach suitable for describing the effective bianisotropic response of a one-dimensional photo-induced THz metamaterial.

\subsubsection{Two-scale approach to the homogenization of a periodic metamaterial}

We here summarize the results of the two-scale homogenization approach developed in Ref. \cite{Ciattoni_1}. Let us consider propagation of a monochromatic electromagnetic field through a periodic media $\varepsilon_r (\textbf{r})=\varepsilon_r \left( \mathbf{r}+\mathbf{\Lambda} \right)$, where $\mathbf{\Lambda}$ is any arbitrary lattice vector. The electric $\textbf{E}$ and magnetic $\textbf{H}$ field amplitudes satisfy Maxwell's equations $\nabla \times \textbf{E} = i \omega \mu_0 \textbf{H}$, $\nabla \times \textbf{H} = -i \omega \varepsilon_0 \varepsilon_r \textbf{E}$, where time dependence $e^{-i\omega t}$ has been assumed. Here, the main assumption is that dielectric permittivity is characterized by a spatial subwavelength modulation and we introduce the parameter $\eta = d/\lambda$ where $d$ is the largest of the lattice basis vector lengths. Considering the long-wavelength limit $\eta \ll 1$, we can develop an asymptotic analysis of electromagnetic propagation. Any field $\mathbf{A}$ ($\mathbf{A}=\mathbf{E},\mathbf{H}$) separately depends on the slow  and fast coordinates ($\mathbf{r}$,  $\mathbf{R} = \mathbf{r} / \eta$, respectively) and $A(\mathbf{r},\mathbf{R})$  can be decomposed as $\mathbf{A} (\mathbf{r},\mathbf{R}) = \overline{\mathbf{A}}(\mathbf{r})  + \tilde{\mathbf{A}}(\mathbf{r},\mathbf{R})$ where the overline denotes the spatial average over the metamaterial unit cell, namely $ \overline{\mathbf{A}}(\mathbf{r}) = \frac{1}{V} \int_C d^3 R \mathbf{A}(\mathbf{r},\mathbf{R})$ ($C$ is the unit cell and $V$ is its volume scaled by $\eta^3$), and the tilde denotes the rapidly varying zero mean residual, i.e. $\tilde{\mathbf{A}} = \mathbf{A} - \overline{\mathbf{A}} $. In our approach, the relative dielectric permittivity only depends on the fast coordinates ($\varepsilon(\mathbf{R}) = \varepsilon_r(\eta \mathbf{R})$) and it can be decomposed as $\varepsilon(\mathbf{R}) = \overline{\varepsilon} + \tilde{\varepsilon}(\mathbf{R})$. Representing each field $\mathbf{A}=\overline{\mathbf{A}}+\tilde{\mathbf{A}}$ as a Taylor expansion up to the first order in $\eta$, we get $\overline{\mathbf{A}} = \overline{\mathbf{A}}_0 (\mathbf{r}) + \overline{\mathbf{A}}_1 (\mathbf{r})  \eta$, $\tilde{\mathbf{A}} = \tilde{\mathbf{A}}_0 (\mathbf{r},\mathbf{R})  + \tilde{\mathbf{A}}_1 (\mathbf{r},\mathbf{R}) \eta $.
From Maxwell's equations, we obtain the averaged equations $\nabla \times \overline{\mathbf{E}} = i \omega \mu_0 \overline{\mathbf{H}}$,
$\nabla \times \overline{\mathbf{H}} = -i \omega \overline{\mathbf{D}}$, where
$\overline{\mathbf{D}} = \varepsilon_0 \Big[ \overline{\varepsilon} \overline{\mathbf{E}} + \overline{ \varepsilon (\tilde{\mathbf{E}}_0   +  \tilde{\mathbf{E}}_1   \eta) } \Big]$.
After some calculations, we obtain $\tilde{\mathbf{E}}_{0} = \hat{\mathbf{e}}_i \left(\partial_i f_j\right) \overline{E}_{0j}$,
$\tilde{\mathbf{E}}_1 = \hat{\mathbf{e}}_i \left[ \left( \partial_i f_j \right) \overline{E}_{1j} + \left(  \delta_{ir} \tilde{f}_j+ \partial_i W_{rj} \right) \frac{\partial \overline{E}_{0j}}{\partial x_r}  \right]$, where the sum is hereafter understood over repeated indices, $\hat{\textbf{e}}_i$ is the unit vector along the $i$-th direction, $\partial_i$ is the partial derivative along $X_i = \hat{\textbf{e}}_i \cdot \mathbf{R}$, $\overline{E}_{0j}= \hat{\textbf{e}}_j \cdot \overline{\mathbf{E}}_{0}$. Here, we have introduced the functions $f_j$ ($\tilde{f}_j$ is the zero mean residual of $f_j$) and $W_{rj}$ satisfying the equations
\begin{eqnarray} \label{f0}
 \nabla_\mathbf{R} \cdot \left( \varepsilon \nabla_\mathbf{R} f_j \right) = -\partial_j \varepsilon, \quad
\nabla_\mathbf{R} \cdot \left( \varepsilon \nabla_\mathbf{R} W_{rj} \right) = -\partial_r \left( \varepsilon \tilde{f}_j\right) - \left( Q_{rj} - \overline{Q_{rj}} \right),
\end{eqnarray}
respectively, where $Q_{rj}=\varepsilon(\delta_{rj}+\partial_r f_j)$ ($\delta_{rj}$ is the Kronecker's delta).
Next, after long calculations \cite{Ciattoni_1}, the effective constitutive relations can be written as
$\overline{D}_i =
\varepsilon_0 \left(
\varepsilon^{(eff)}_{ij} \overline{E}_j +
\alpha^{(eff)}_{ijr} \frac{\partial \overline{E}_j}{\partial x_r} \right)$,
$\overline{B}_i = \mu_0 \overline{H}_i$, where
$\varepsilon^{(eff)}_{ij} = \frac{1}{2} \overline{  \left( Q_{ij} + Q_{ji}  \right)}$ and
$\alpha^{(eff)}_{ijr} =  \eta \overline{ \left( Q_{ri} \tilde{f}_j - Q_{rj} \tilde{f}_i \right) }$.
Exploiting the Serdyukov-Fedorov transformation, the constitutive relations can  be transformed to a symmetric form \cite{Serdy}, i.e.
\begin{eqnarray} \label{constitutive1}
\overline{\textbf{D}'} =  \varepsilon_0 \varepsilon'^{(eff)} \overline{\textbf{E}'} - \frac{i}{c} \kappa^{(eff)T} \overline{\textbf{H}'},  \quad
\overline{\textbf{B}'} = \frac{i}{c} \kappa^{(eff)}  \overline{\textbf{E}'} + \mu_0 \overline{\textbf{H}'},
\end{eqnarray}
where
\begin{eqnarray} \label{eff_last}
\varepsilon'^{(eff)}_{ij} = \varepsilon^{(eff)}_{ij} + \kappa^{(eff)}_{ir} \kappa^{(eff)}_{rj}, \quad
\kappa^{(eff)}_{ij} = \eta k_0 \left[ \epsilon_{imj} \overline{\varepsilon f_m} +\left(\epsilon_{imn} \delta_{jq} + \frac{1}{2} \epsilon_{mqn} \delta_{ij} \right) \overline{\varepsilon f_m \partial_q f_n}\right],
\end{eqnarray}
where $\epsilon_{imn}$ is the Levi-Civita symbol. $\kappa_{ij}^{(eff)}$ is the effective chiral medium tensor and it is provided by the first order spatial dispersion.

\subsubsection{Extended Landau-Lifshitz-Looyenga Effective-Medium Approach}

Here, we summarize the derivation on a homogenization approach, developed in Ref\cite{Rizza_L}, which can be considered the extension of Landau-Lifshitz-Looyenga (LLL) effective-medium approach in the context of periodic metamaterials. The LLL approach is an effective medium theory for evaluating the electrostatic effective dielectric permittivity of an isotropic mixture in the case where the dielectric contrast is low \cite{Landau}. Specifically we use the approach summarized in the above section in the situation where
\begin{equation} \label{low}
\varepsilon = \overline{\varepsilon}+ \tilde{\varepsilon}({\bf R}) \equiv \overline{\varepsilon}+ \tau {  \Delta \varepsilon}({\bf R}),
\end{equation}
where $\tau \ll 1$. Accordingly, we expand the potential field ${\bf f} = f_j \hat{\bf e}_j$ as a power series in the small parameter $\tau$ up to the second order
\begin{equation} \label{low2}
{\bf f}= {\bf f}^{(0)}+\tau {\bf f}^{(1)}+ \tau^2 {\bf f}^{(2)}.
\end{equation}
Substituting Eq.(\ref{low}) and Eq.(\ref{low2}) into the first of Eq.(\ref{f0}) and extracting equations for each order in $\tau$, we get
\begin{eqnarray} \label{low3}
\nabla_{\bf R}^2 f_j^{(0)}&=&0, \nonumber \\
\overline{\varepsilon} \nabla_{\bf R}^2 f_j^{(1)}&=& -\partial_j \Delta \varepsilon - \nabla_{\bf R} \cdot ({  \Delta \varepsilon} \nabla_{\bf R} f^{(0)}_j  ) , \nonumber \\
\overline{\varepsilon}  \nabla_{\bf R}^2 f_j^{(2)}&=&- \nabla_{\bf R} \cdot ({  \Delta \varepsilon} \nabla_{\bf R} f^{(1)}_j  ).
\end{eqnarray}
In order to solve such differential equations, we consider the Fourier series for ${ \Delta \varepsilon}$ and for the fields $f_j^{(m)}$ ($m=0,1,2$) which are given by, respectively,
\begin{eqnarray} \label{low4}
{\bf  \Delta \varepsilon}&=&\sum_{{\bf G} \neq {\bf 0}} {  \Delta \varepsilon}_{\bf G} e^{i {\bf G} \cdot {\bf R}}, \nonumber \\
{\bf f}^{(m)}&=&\sum_{\bf G} {\bf f}^{(m)}_{\bf G} e^{i {\bf G} \cdot {\bf R}}.
\end{eqnarray}
Using the Fourier series of Eqs.(\ref{low4}), we obtain an explicit solution of the set of Eqs.(\ref{low3}) which reads
\begin{eqnarray} \label{fff}
f_j= i \tau \sum_{{\bf G} \neq {\bf 0}} \left( \frac{{ \Delta \varepsilon}_{\bf G}}{\overline{\varepsilon}} \frac{G_j}{|{\bf G}|^2}- \tau                   \sum_{{ {\bf G}'}\neq {\bf 0}} \frac{{ \Delta \varepsilon}_{{\bf G}'} { \Delta \varepsilon}_{{\bf G}-{ {\bf G}'}} }
{\overline{\varepsilon}^2} \frac{({\bf G} \cdot { {\bf G}'}) {G_j'} }{|{\bf G}|^2 |{ {\bf G}'}|^2}  \right) e^{i {\bf G} \cdot {\bf R}}.
\end{eqnarray}
Substituting Eq.(\ref{fff}) into Eqs.(\ref{eff_last}), the analytical expression for the electromagnetic dielectric and chiral tensors are given by, respectively,
\begin{eqnarray} \label{eps_kappa}
\varepsilon'^{(eff)}_{ij} & = & \overline{\varepsilon} \delta_{ij}-\frac{\tau^2}{\overline{\varepsilon}}
\sum_{{\bf G} \neq {\bf 0}}  \frac{{ \Delta \varepsilon}_{- \bf G} G_i} {|{\bf G}|^2}\Bigg[  { \Delta \varepsilon}_{\bf G} G_j
- \tau \sum_{{ {\bf G}'} \neq {\bf 0}} \frac{  { \Delta \varepsilon}_{\bf K} { \Delta \varepsilon}_{{\bf G}-{ {\bf G}'}}}{\overline{\varepsilon}}  \frac{({\bf G} \cdot { {\bf G}'}) {G_j'}} {|{ {\bf G}'}|^2} \Bigg],
\nonumber \\
\kappa_{ij}^{(eff)}& = & i \eta k_0 \frac{\tau^3}{\overline{\varepsilon}^2} \sum_{{\bf G} \neq {\bf 0}, { {\bf G}'}\neq {\bf 0}}
\frac{{ \Delta \varepsilon}_{-{\bf G}-{ {\bf G}'}} {\Delta \varepsilon}_{{\bf G}} { \Delta \varepsilon}_{{ {\bf G}'}} }{|{\bf G}|^2 |{ {\bf G}'}|^2} \times \nonumber \\
&&  \Bigg[ ({\bf G} \cdot { {\bf G}'}) \epsilon_{imj} { {G}'_m} + \left( 1+ 2 \frac{{\bf G} \cdot { {\bf G}'} }{|{\bf G}|^2}  \right) \left( \epsilon_{imn} \delta_{jq}+\frac{1}{2} \epsilon_{mqn} \delta_{ij} \right)
G_q G_m { {G}'_n} \Bigg],
\end{eqnarray}
where we have neglected the fourth and higher order terms in $\tau$.

\subsubsection{Analytical effective medium theory for all-photodesigned one-dimensional THz metamaterials}

Here, by using Eqs.(\ref{eps_kappa}), we obtain analytical expressions for the dielectric and chiral tensors of an all-photodesigned 1D THz metamaterial. In the simplified situation considered in the Sec.IV of the main manuscript (where we neglect the IR propagation within the thin slab together with the nonlinear contributions to the carrier dynamics and the $z$-dependence of the carrier density $N$), we get the THz microscopic dielectric contrast
\begin{eqnarray}
\label{perm_THz0}
\varepsilon^{(THz)} &= &\varepsilon_b + g N,
\end{eqnarray}
where
\begin{equation} \label{NN0}
N = \beta I_0 \left\{ {1 + d_1 \cos \left( {\frac{{2\pi }}{\Lambda }x} \right) + \chi d_2 \sin \left( {\frac{{4\pi }}{\Lambda }x} \right)} \right\}.
\end{equation}
In the Eq.(\ref{perm_THz0}) and Eq.(\ref{NN0}), $g = - e^2/ \left(\varepsilon_0 m^* \right) \left[\omega^{(THz)} \left( \omega^{(THz)}+i \gamma_D \right) \right]^{-1}$, $\varepsilon_b$ is the THz background dielectric constant in the absence of the optical beam, $-e$ is the electron charge unit, $m^*$ is the electron effective mass and $\gamma_D$ is the free electron relaxation rate; $\beta  = \frac{{{\mathop{\rm Im}\nolimits} \left( {\varepsilon^{(IR)}(0) } \right)}}{{A \hbar c}}$, $d_1  = \frac{1}{2}\left[ {1 + \left( {2\pi \frac{{L_D }}{\Lambda }} \right)^2 } \right]^{ - 1}$, $d_2  = \frac{1}{2}\left[ {1 + \left( {4\pi \frac{{L_D }}{\Lambda }} \right)^2 } \right]^{ - 1}$, $Im \left( \varepsilon^{(IR)}(0) \right) = 0.14$.

Comparing Eq.(\ref{perm_THz0}) with Eq.(\ref{low}) we get $\bar \varepsilon  = \varepsilon _b  + g \beta I_0$ and $\Delta \varepsilon^{(THz)} = \tau \Delta \varepsilon = g \left( N - \beta I_0 \right)$ so that Eqs.(\ref{eps_kappa}) yield $\varepsilon'^{(eff)} = \textrm{diag}(\varepsilon _\parallel,\varepsilon _ \perp,\varepsilon _ \perp)$ and $\kappa^{(eff)}_{ij}=\varepsilon_{ij1} \kappa$, where
\begin{eqnarray} 
\varepsilon _\parallel   &=&
 \bar \varepsilon - \frac{1}{4} \left[ d_1^2 + \left( d_2 \chi  \right)^2  \right]  \frac{\left(g \beta  I_0 \right)^2 }{\bar \varepsilon },
\quad \quad \varepsilon _ \perp   = \bar \varepsilon, \nonumber \\
\kappa  &=& \eta \chi  \frac{3  d_1^2 d_2}{8} \frac{\left({ {g\beta  I_0 } }\right)^3}{{\bar \varepsilon}^2},
\end{eqnarray}

\end{document}